\documentclass[review]{elsarticle}

\usepackage{lineno,hyperref,xcolor}
\modulolinenumbers[5]

\journal{Journal of \LaTeX\ Templates}

\begin{document}

\begin{frontmatter}

\title{High dynamic range base design and characterization of an 8-inch photomultiplier tube CR365-02-2 for the LHAASO-MD experiment}
%\title{Characterization of the CR365-02-2 8-inch photomultiplier tube for the LHAASO-MD experiment with a high dynamic range base}
%\title{Elsevier \LaTeX\ template\tnoteref{mytitlenote}}


\author[add1,add2]{Yang Li} 
\author[add1,add2]{Kun Jiang\corref{cor1}}
\cortext[cor1]{Corresponding author}
\ead{kjiang@ustc.edu.cn}
\author[add1,add2]{Xiaokun Zhao}
\author[add1,add2]{Xin Li}
\author[add1,add2]{Cheng Li}
\author[add1,add2]{Zebo Tang\corref{cor1}}
\ead{zbtang@ustc.edu.cn}
%\author[add1,add2]{Zehua Cao} 
%\author[add1,add2]{Cheng Li}
%\author[add1,add2]{Yang Li} 
%\author[add1,add2]{Ziwei Li} 
%\author[add1,add2]{Ziyang Li} 
%\author[add1,add2]{Zheng Liang}
%\author[add1,add2]{Pengzhong Lu} 
%\author[add1,add2]{Kaifeng Shen}
%\author[add1,add2]{Kaiyang Wang} 
%\author[add1,add2]{Yan Wang} 
%\author[add1,add2]{Xin Wu} 
\address[add1]{State Key Laboratory of Particle Detection and Electronics, University of Science and Technology of China, Hefei 230026, China}
\address[add2]{Department of Modern Physics, University of Science and Technology of China, Hefei 230026, China}

\begin{abstract}
Ultra-high-energy ($>$ 100 TeV) gamma-ray detection benefits from the use of muon detectors (MDs) thanks to their capability of suppressing the cosmic-ray background. More than 1100 8-inch photomultiplier tubes (PMTs), CR365-02-2 from Beijing Hamamatsu Photon Techniques INC. (BHP), will be deployed for the LHAASO-MD experiment. In this paper, the design of the photomultiplier base for CR365-02-2 is presented. Signals are extracted from two outputs: the anode and the 7th dynode. This design ensures a high dynamic range, from single photoelectron (SPE, peak-to-valley ratio $>$ 2), up to bunches of $7\times10^5$ photoelectrons, with an equivalent anode peak current of up to 1600 mA. The anode-to-dynode amplitude ratio (A/D) is below 160 in order to ensure enough overlap between the two outputs. Test results of 4 PMTs, including the SPE response, A/D, linearity, dark count rate, and afterpulse rate, are reported. 
\end{abstract}


\end{frontmatter}

%\linenumbers

% main text

%\paragraph{Functionality} The Elsevier article class is based on the standard article class and supports almost all of the functionality of that class. In addition, it features commands and options to format the
%\begin{itemize}
%\item document style
%\item baselineskip
%\end{itemize}

%\begin{enumerate}[(1)]
%\item Group the authors per affiliation.
%\item Use footnotes to indicate the affiliations.
%\end{enumerate}

\section{Introduction}\label{intro}
The cosmic-ray factories that accelerate particles up to $10^{15}$ eV (PeV) are called PeVatrons. To confirm the Galactic origin of PeV cosmic rays, we need to identify PeVatrons in our Galaxy~\cite{Aloiso:2018hbl, HAWC:2019tcx, TibetASgamma:2021tpz}. Cosmic rays, being charged, are deflected by the magnetic fields present in the interstellar space, and are difficult to trace back to their sources. However, high energy neutral gamma-rays can be used to identify PeVatrons. Two scenarios have been proposed to explain the high energy gamma-ray production. One is the hadronic scenario. Neutral pions, created when cosmic rays interact with the environment, decay to two gamma-rays~\cite{Naito:1994xr}. The other is the leptonic scenario via the inverse Compton scattering of relativistic electrons~\cite{Jones:1968zza, Blumenthal:1970gc}. In both cases, the principal signature is a hard gamma-ray energy spectrum that extends beyond 100 TeV without any cut-off.

The Large High Altitude Air Shower Observatory (LHAASO)~\cite{Cao:2010zz} is a multipurpose project for high energy gamma-rays and cosmic-rays detection in Sichuan Province (PR China) at an altitude of 4410 meters. It consists of an extensive air shower array covering 1 km$^2$~(KM2A), a 78,000 m$^2$ water Cherenkov detector array (WCDA), and 18 wide-field air Cherenkov/fluorescence telescopes (WFCTA)~\cite{He2018Design}. KM2A focuses on searching for gamma-ray sources of energy above 30 TeV and on measuring primary cosmic rays at an energy between 10 TeV and 100 PeV~\cite{Cui:2014bda}. It is composed of 5195 electromagnetic particle detectors (EDs) and 1188 muon detectors (MDs). An ED is a type of plastic scintillator detector, and is designed to measure the density and arrival times of secondary particles in extensive air showers (EAS)~\cite{Jia:2014kqa, Jing:2014loa, Lv:2015zba, Zhang:2017cor}. The MDs are designed to measure the shower muon content~\cite{Zuo:2015qva, LHAASO:2018wcf}. Since gamma showers are muon-poor, measuring the numbers of electromagnetic particles and muons in EAS can effectively discriminate between primary gamma-rays and nuclei. Using data from the large scale MD array, it is possible to reduce the cosmic-ray background in the KM2A by factors of 1000 and 500000 at 50 TeV and 1 PeV, respectively~\cite{cao2021peta}. This excellent background rejection power plays an important role in the detection of ultra-high-energy photons, up to 1.4 PeV, from 12 gamma-ray Galactic sources~\cite{cao2021ultrahigh}.

MDs are placed in a triangular grid with a 30 m spacing. To accurately measure the number of muons, an MD should have sufficiently good detection efficiency ($>95\%$) and signal charge resolution ($<25\%$). The purity of muons detected in hadronic showers should be $>$ 95\%. The dynamic range of the MD is 1--$10^4$ muons, with the upper limit corresponding to the number of muons near the shower core of the highest energy (up to 100 PeV) hadrons to be detected~\cite{Zuo:2015qva,He2018Design}. The time resolution required is only about 10 ns, because MDs are not used to trigger or reconstruct EAS.

\begin{figure}[htbp]
\begin{center}
\includegraphics[keepaspectratio,width=0.98\textwidth]{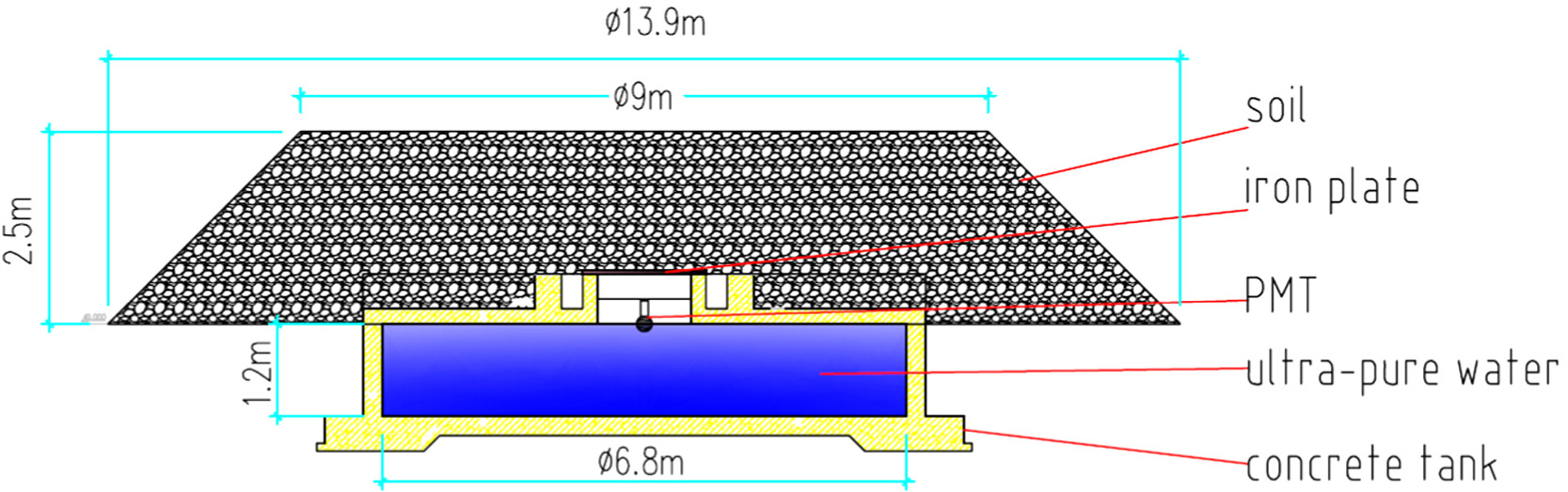}
\vspace*{+0mm}
\caption{Schematic of LHAASO muon detector~\cite{Zuo:2015qva}.} \label{fig:md}
\end{center}
\end{figure}

For the LHAASO-MD, the design of a water Cherenkov detector underneath the soil was chosen (Fig.~\ref{fig:md}). A bag is used to enclose pure water. The inner layer of the water bag is made of Tyvek 1082D (DuPont), an opaque material with high reflectivity for near-UV lights. The water bag is held by a cylindrical concrete tank with an inner diameter of 6.8 m and a height of 1.2 m. An 8-inch photomultiplier tube (PMT) is installed at the tank roof centre, pointing downward through a highly transparent window into the water. The tank is covered by a steel lib and overburden soil with a thickness of 2.5 m, which shield most of the electromagnetic component of the showers. 

The number of photoelectrons (PEs) collected by the PMT, after reaching its maximum, decreases exponentially over time because of the absorption of water and reflection of the Tyvek liner. The decay time of a single muon signal is approximately 200 ns~\cite{Zuo:2015qva}. The arrival time dispersion of PEs is much larger than the intrinsic time response of the PMT. The average response of the PMT to single muon signals is about 70 PEs, corresponding to a peak current of 0.16 mA at a PMT gain of $2\times 10^6$~\cite{Zuo:2015qva}. Therefore, the PMT dynamic range for MD units must be from single PE up to bunches of $7\times 10^5$ PEs, corresponding to a peak current of 1600 mA. The specifications for the 8-inch PMTs are listed in Table~\ref{tab:cut}. After evaluating various types of PMTs, the CR365-02-2 from Beijing Hamamatsu Photon Techniques INC. (BHP) was chosen for the LHAASO-MD. The CR365-02-2 has the same structure as the CR365-02-1 used for LHAASO-WCDA~\cite{Jiang:2020rdv}. It has 10 dynodes, features a large photocathode, and is characterized by a good single photoelectron (SPE) resolution, low dark noise rate, and low afterpulse rate.
\begin{table*}
\centering
  \begin{tabular}{c|c}
    Parameters  &  Specifications  \\\hline
    Working gain &  $2\times10^6$\\
    Peak-to-valley ratio   &  $>$ 2.0  \\
    Charge resolution for single muon   &  $<$ 25\%  \\
    Anode linearity (within 5\%)    &    0--25 mA\\
    7th dynode linearity (within 5\%, equiv. anode)    &   1600 mA\\
    Anode-to-dynode amplitude ratio  & 80--160 \\
    Dark count rate ($>1/3$ PE)   &   $<$ 5000 cps \\
    Afterpulse rate (100--10000 ns)   &   $<$ 5\% \\
\end{tabular}
\caption{Specification requirements of CR365-02-2.}\label{tab:cut}
\end{table*}

In section~\ref{basedesign} of this paper, the design of a high dynamic range readout base for the CR365-02-2 of the LHAASO-MD is described. The performances, including SPE response, anode-to-dynode amplitude ratio (A/D), linearity within the dynamic range, dark count rate, and afterpulse rate, will be discussed in section~\ref{measurements}. 

\section{The high dynamic range base design}\label{basedesign}
\begin{figure}[htbp]
\begin{center}
\includegraphics[keepaspectratio,width=0.98\textwidth]{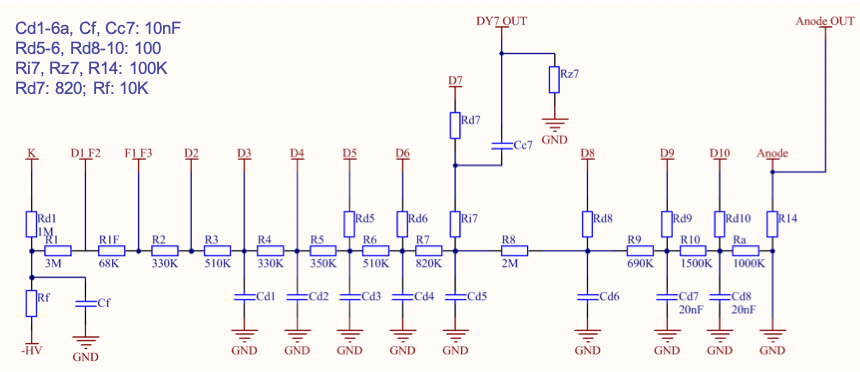}
\vspace*{+0mm}
\caption{The base circuit used for the LHAASO-MD.} \label{fig:base}
\end{center}
\end{figure}

The schematic of the base circuit is shown in Fig.~\ref{fig:base}. It is made only from resistors and capacitors. 
%The total resistance is 11.12 M$\Omega$. A negative high voltage is supplied since a highly transparent window separates the PMT from the water. 
To guarantee a high dynamic range, signals are extracted from two outputs: one from the anode and the other from the 7th dynode (DY7). The signal from DY7 will be used when the anode signal starts to deviate from the linear response. Since the space charge effect between the 6th and 7th dynodes is less than that between the last dynode and the anode, reading signals from DY7 can significantly extend the dynamic range of the PMT. Another advantage of using two outputs with different gains is to lower the requirement on the dynamic range of the readout electronics. With A/D slightly below 160, it is estimated that the dynamic range of readout electronics can be reduced by a factor of 40. The A/D must be less than 160 to ensure a large enough overlap region between the two outputs. This requirement increases the difficulties to reach the high dynamic range required by the experiment.

Three methods are further applied to improve the linearity of the anode and dynode outputs. First, the voltage division ratios between the last few dynodes and the anode are suitably enlarged to overcome the space charge effects. Second, two isolation resistors (Ri7 and R14) isolate the bleeder from the signal outputs~\cite{Genolini:2003vf}. The high-value isolation resistors (100 k$\Omega$ each) will allow only a tiny part of the signal current to flow into the bleeder. Third, the decoupling capacitors (Cd1--Cd8) are used in parallel to evacuate the charge during the occurrence of large pulses, and thus keep the voltage of the last few stages stable. Damping resistors (Rd5--Rd10) on the last dynodes are used to reduce the ringing effect. Oscillations on the falling edge of DY7, due to parasitic inductances of PMT inner leads, are significantly reduced by Rd7 (820 $\Omega$). The filter resistor Rf and capacitor Cf are used to stabilize the high voltage supply.  

\section{Measurements and results}\label{measurements}
\subsection{Single photoelectron response}
\begin{figure}[htbp]
\begin{center}
\includegraphics[keepaspectratio,width=0.68\textwidth]{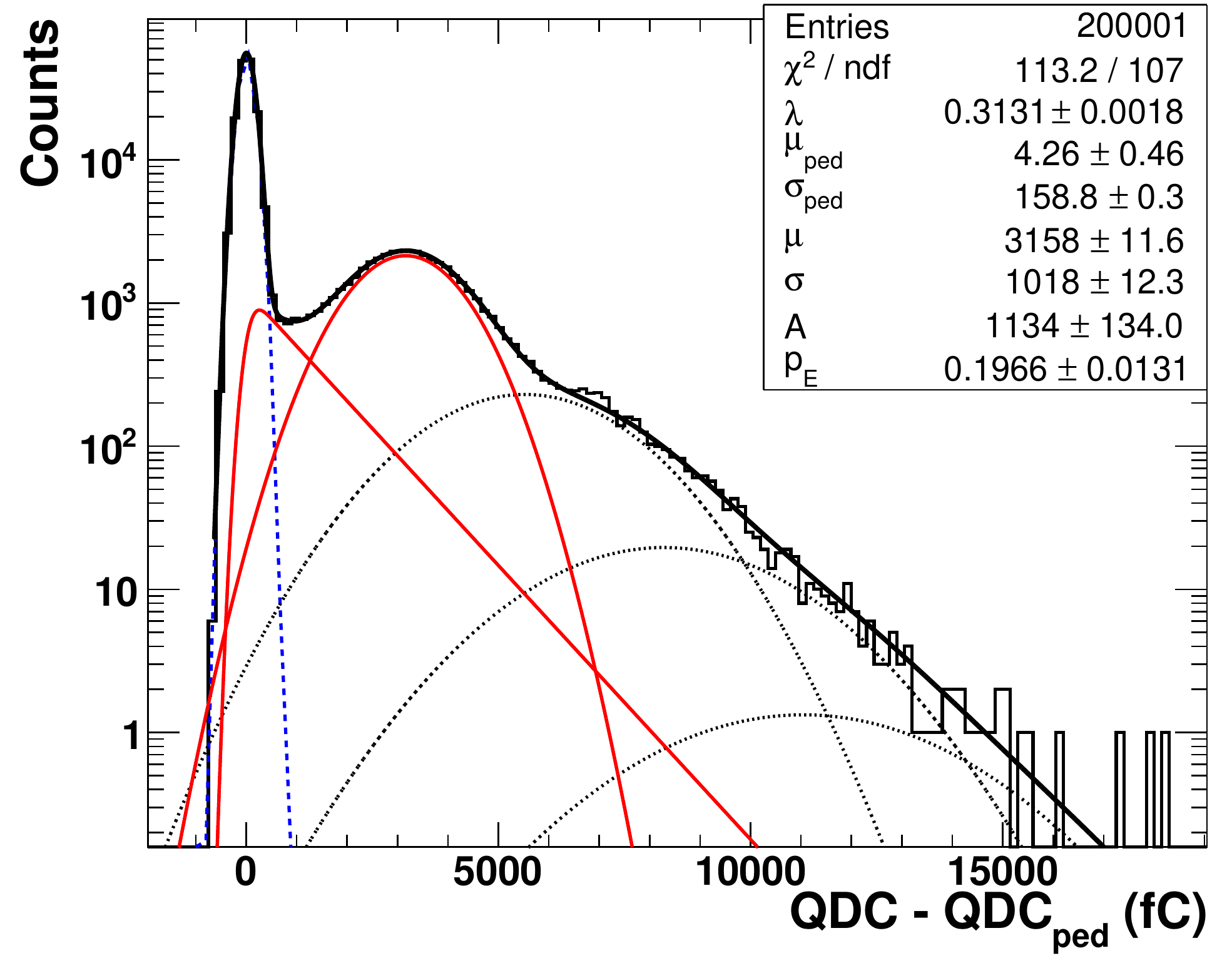}
\caption{The SPE charge spectrum at a gain of $2\times10^6$. The blue dashed line is the pedestal. The red lines are the Gaussian and the exponential parts of the SPE response. The black dotted lines correspond to 2, 3, and 4 PEs} \label{fig:spe}
\end{center}
\end{figure}

The experimental setup used to obtain the results shown here is similar to the one to test the PMTs for LHAASO-WCDA and is described elsewhere~\cite{Jiang:2020rdv}. The PMT is illuminated by a pico-second laser pulser (Hamamatsu PLP-10). To have enough statistics, the light intensity of the laser is adjusted so that the mean number of PEs per trigger is about 0.3. Fig.~\ref{fig:spe} shows the anode SPE charge spectrum at 1260 V, corresponding to a gain of $2\times10^6$, with a 10 times fast amplifier (CAEN N979). The  ideal SPE response of a PMT is modelled as a sum of a Gaussian and an exponential~\cite{Dossi:1998zn}:
\begin{eqnarray}
	S_{0}\left(x\right)=\frac{p_E}{A}e^{-\frac{x}{A}} + \frac{1-p_E}{g_N}\frac{1}{\sqrt{2\pi}\sigma}e^{-\frac{\left(x-\mu\right)^{2}}{2\sigma^2}}, x>0
\label{eq:S0}
\end{eqnarray}
where $x$ is the variable charge,  $A$ is the slope of the exponential part of the $S_{0}(x)$, $p_E$ is the fraction of events under the exponential function, $\mu$ is the mean of the Gaussian part of the SPE response which is used to define the PMT gain~\cite{IceCube:2010dpc}, $\sigma$ is the standard deviation of the Gaussian part of the SPE response. The factor $g_N$ compensates for the usage of a truncated Gaussian. 
\begin{eqnarray}
	g_N=\frac{1}{2}\left(1+ \textrm{Erf}\frac{\mu}{\sqrt2\sigma}\right)\;,
\label{eq:gN}
\end{eqnarray}
where $\textrm{Erf}(x)$ is the error function. The realistic PMT spectrum can be described by~\cite{Dossi:1998zn}: 
\begin{eqnarray}
f(x) =e^{-\lambda}\lambda \left[\frac{p_E}{2A}e^{\frac{\sigma_{ped}^2-2A\left(x-\mu_{ped}\right)}{2A^2}}\left(1+ \textrm{Erf}\frac{A(x-\mu_{ped})-\sigma_{ped}^2}{\sqrt2A\sigma_{ped}}\right)+\frac{1-p_E}{g_N}\frac{1}{\sqrt{2\pi(\sigma^2+\sigma_{ped}^2)}}e^{-\frac{\left(x-\mu-\mu_{ped}\right)^{2}}{2(\sigma^2+\sigma_{ped}^2)}} \right] \nonumber \\
+ \frac{e^{-\lambda}}{\sqrt{2\pi}\sigma_{ped}}e^{-\frac{\left(x-\mu_{ped}\right)^{2}}{2\sigma_{ped}^2}} + \sum^4_{n=2}\frac{e^{-\lambda}\lambda^n}{n!}\frac{1}{\sqrt{2\pi(n\sigma_1^2+\sigma_{ped}^2)}}e^{-\frac{\left(x-n\mu_1-\mu_{ped}\right)^{2}}{2(n\sigma_1^2+\sigma_{ped}^2)}}
\label{eq:fx}
\end{eqnarray}
where $\lambda$ is the mean number of PEs collected by the first dynode, 
$\mu_{ped}$  and $\sigma_{ped}$ are the mean and the standard deviation of the pedestal, respectively. $\mu_1$  and $\sigma_1$ are the mean and the standard deviation of the SPE response.
\begin{equation}
	\mu_1 \approx (1-p_E)\mu+p_EA
\label{eq:mu1}
\end{equation}
\begin{equation}
	\sigma_{1}^2 \approx (1-p_E)(\sigma^2+\mu^2)+2p_EA^2-\mu_1^2
\label{eq:sigma1}
\end{equation}

The peak-to-valley ratio (P/V), defined as the maximum of the SPE peak over the minimum of the valley between the pedestal and the SPE peak, is 3.5. The SPE energy resolution ($\sigma_1/\mu_1$) is 47.7\%. Since the average response of the PMT to single vertical muons ($N_{PE}$) is about 70 PEs~\cite{Zuo:2015qva}, the charge resolution of the PMT for a single muon is: 
\begin{equation}
\frac{47.7\%}{\sqrt{N_{PE}}}\otimes\frac{1}{\sqrt{N_{PE}}}=\sqrt{\left(\frac{47.7\%}{\sqrt{70}}\right)^2+\left(\frac{1}{\sqrt{70}}\right)^2}=13.2\%,
\label{eq:resSingleMuon}
\end{equation}
The P/V and the energy resolution meet the requirements of LHAASO-MD. 

\subsection{Non-linearity and A/D}
The MDs do not participate in the triggering or reconstruction of the direction of a shower. Cherenkov photons, produced by air shower muons crossing water, can be reflected by the Tyvek liner of the water bag. The trailing edge of the single muon waveform can be described by an exponential function, with a decay time of about 200 ns~\cite{Zuo:2015qva}. Since the arrival time dispersion of PEs is much larger than the intrinsic time response of the PMT, the dynamic range is determined by the peak current. 

\begin{figure}[htbp]
\begin{center}
\includegraphics[keepaspectratio,width=0.68\textwidth]{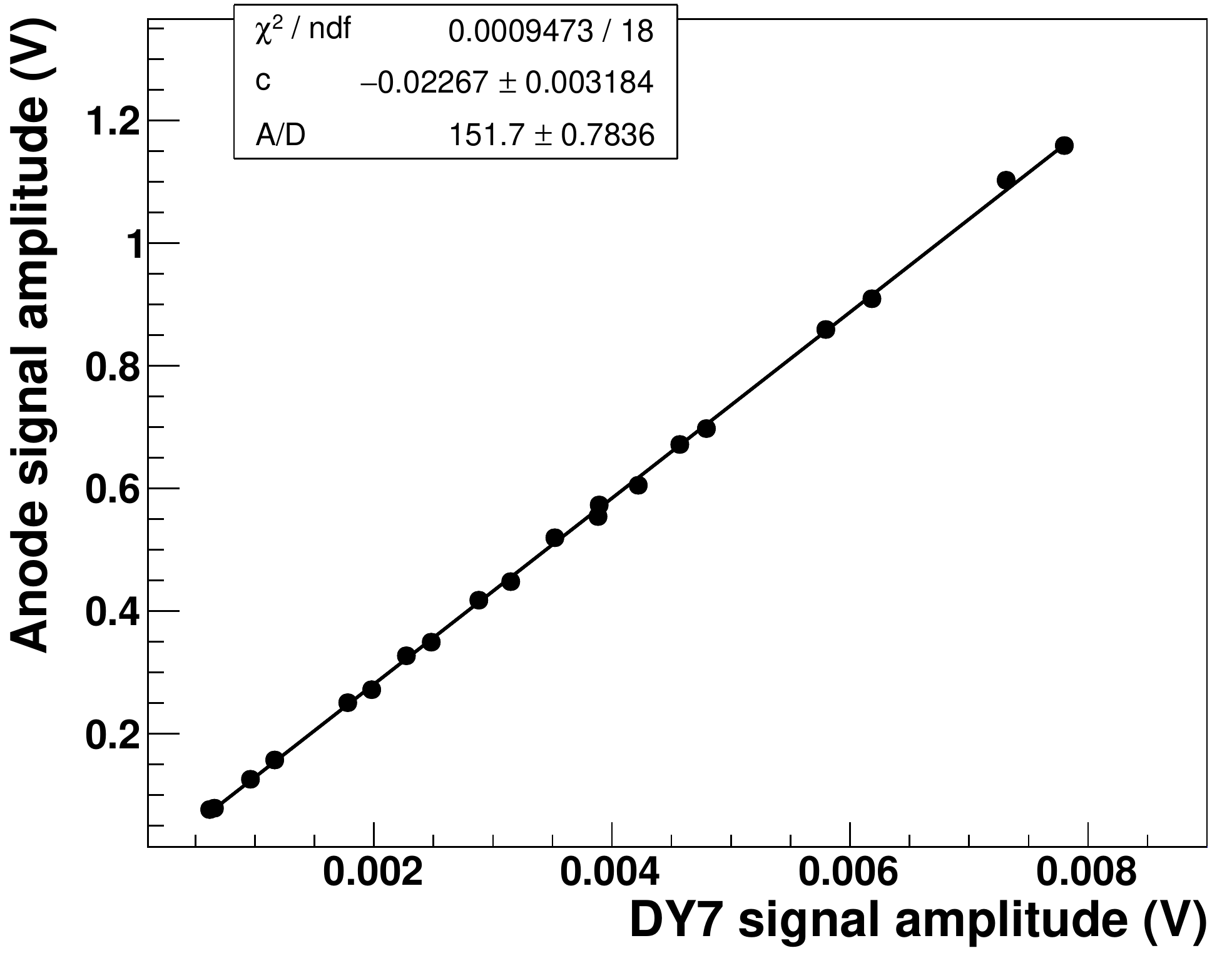}
\caption{The amplitude correlation between the anode and the DY7.} \label{fig:ad}
\end{center}
\end{figure}

The waveforms of the PMT outputs are recorded by a Digitizer (CAEN V1743). Two blue LEDs (Hebei 510LB7C), driven by a pulse generator (Tektronix AFG3252), were used to illuminate the PMTs. The PMTs were tested using three kinds of LED driving pulses: rectangular pulses with a width of 100 ns and 200 ns, and triangular pulses with a 400 ns duration. The rise time and fall time of the pulses were 2.5 ns. The non-linearity and A/D show no difference under these three kinds of pulses~\cite{XiaokunThesis}. Therefore 100 ns rectangular driving pulses were chosen for the following measurements. Fig.~\ref{fig:ad} shows the amplitude correlation between the signals recorded at the anode and the DY7. The A/D, which is the slope of the curve, is $151.7\pm0.8$. The specification for A/D ($<160$) ensures enough overlaps between the two outputs. 

\begin{figure}[htbp]
\begin{center}
\includegraphics[keepaspectratio,width=0.49\textwidth]{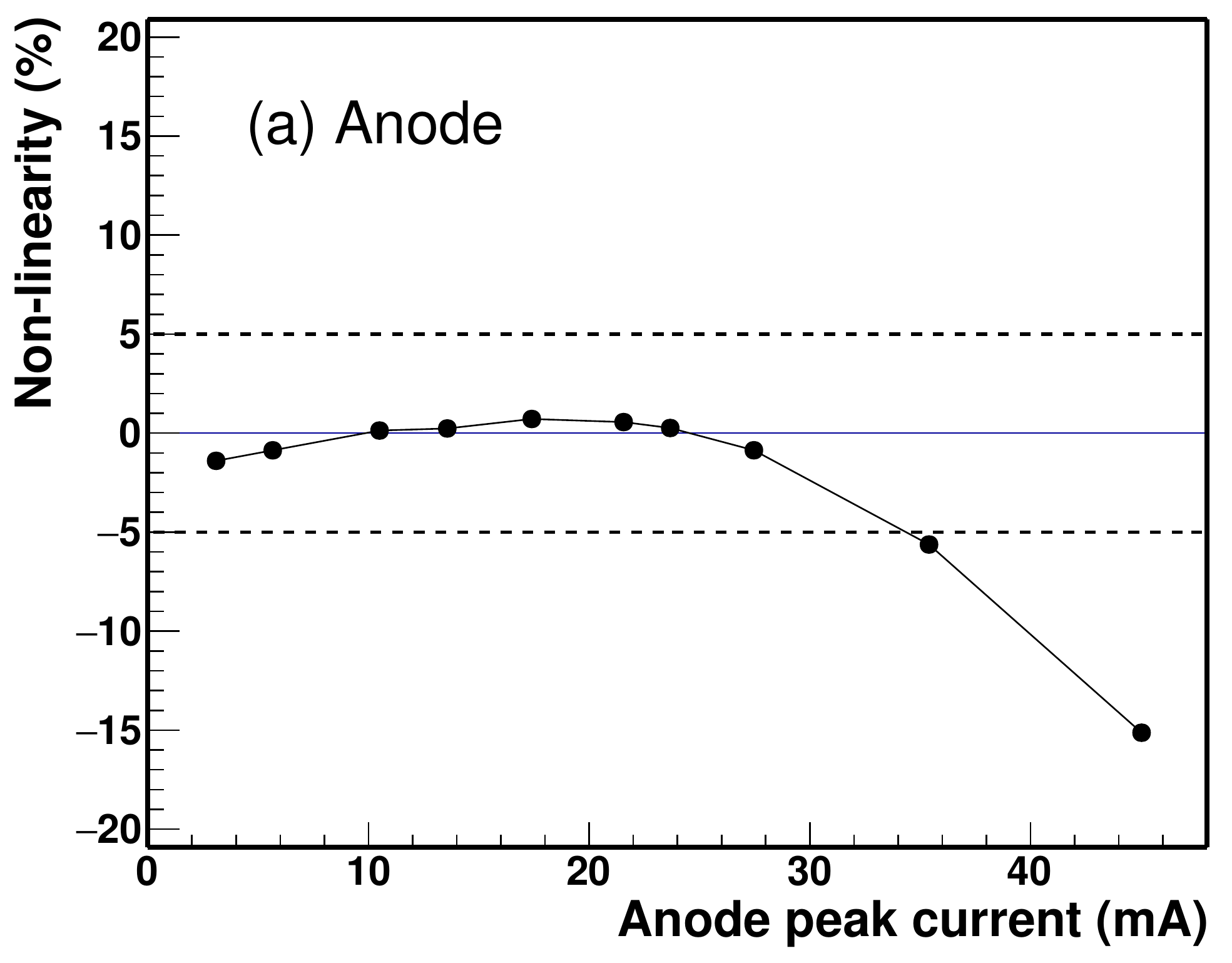}
\includegraphics[keepaspectratio,width=0.49\textwidth]{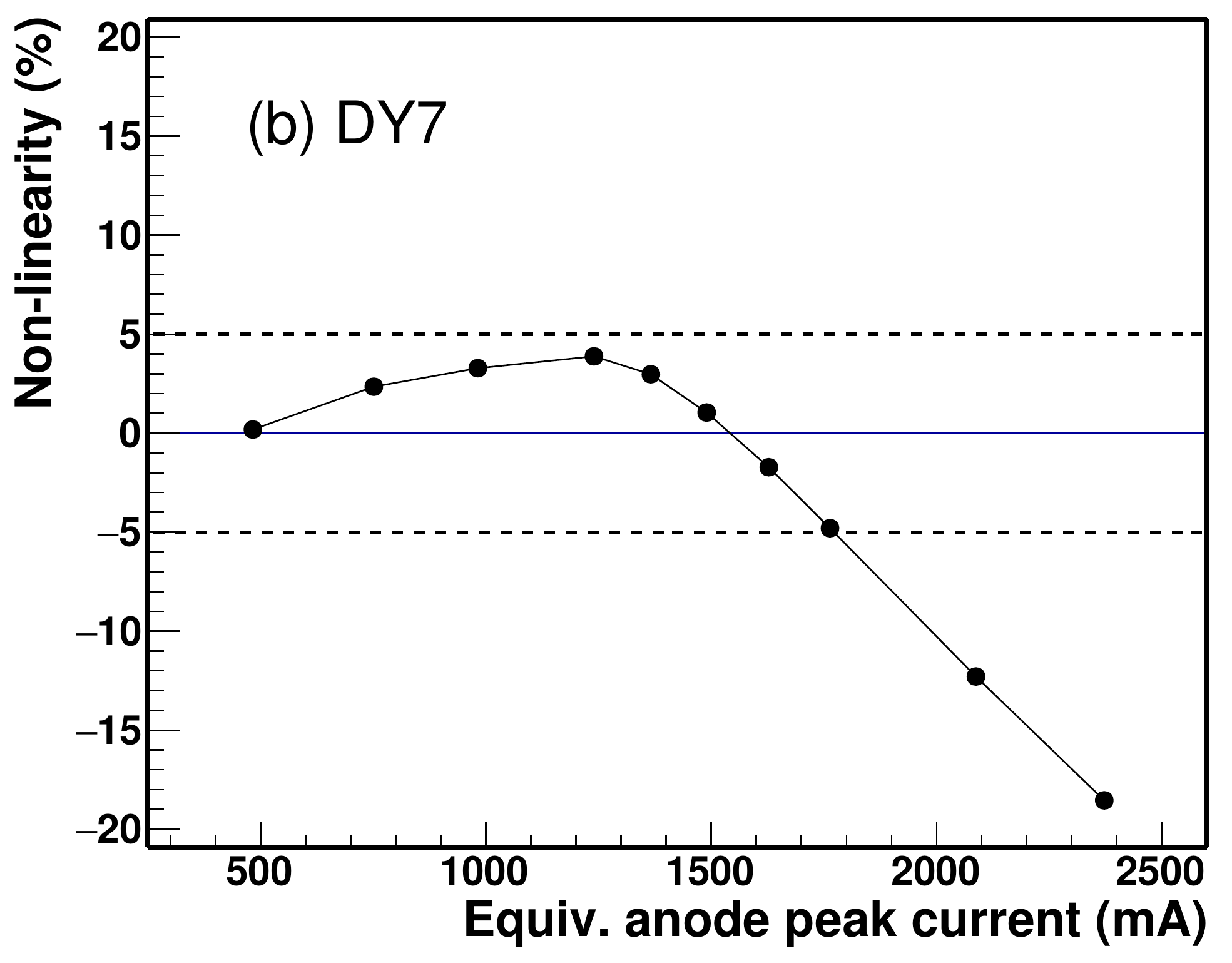}
\caption{(a) Non-linearity of the anode as a function of anode peak current. (b) Non-linearity of DY7 as a function of equivalent anode peak current.} \label{fig:nl}
\end{center}
\end{figure}

The A-B method is used to determine the non-linearity~\cite{Barnhill:2008zz,Jiang:2020rdv}. The dynamic range is defined as where the deviation from the ideal linear current reaches -5\%. Fig.~\ref{fig:nl} shows non-linearity curves for the anode and the DY7 as a function of the peak current. The peak current of the anode and the DY7 are 34.4 mA and 11.7 mA, respectively, when the non-linearity is -5\%. The A/D is used to convert the peak current of DY7 to an equivalent anode peak current. With an A/D of 151.7 measured above, the equivalent anode peak current is 151.7 $\times$ 11.7 mA = 1775 mA, which meets the specification for DY7 (equivalent anode peak current $>$ 1600 mA). 

\subsection{Dark count rate and afterpulse rate}
Dark noise and afterpulses are the primary sources of undesired background signals for PMTs. They mimic real physical signals, increase the probability of a random trigger, and shift the energy scale of the detector. In the region with medium supply voltage, the dark noise is dominated by thermal electron emission~\cite{pmthandbookv4e}. Before measuring the dark count rate, the PMT was put in a dark box for more than 6 h. With a threshold of 1/3 PE, the measured dark count rate was 505 cps. 
\begin{figure}[htbp]
\begin{center}
\includegraphics[keepaspectratio,width=0.66\textwidth]{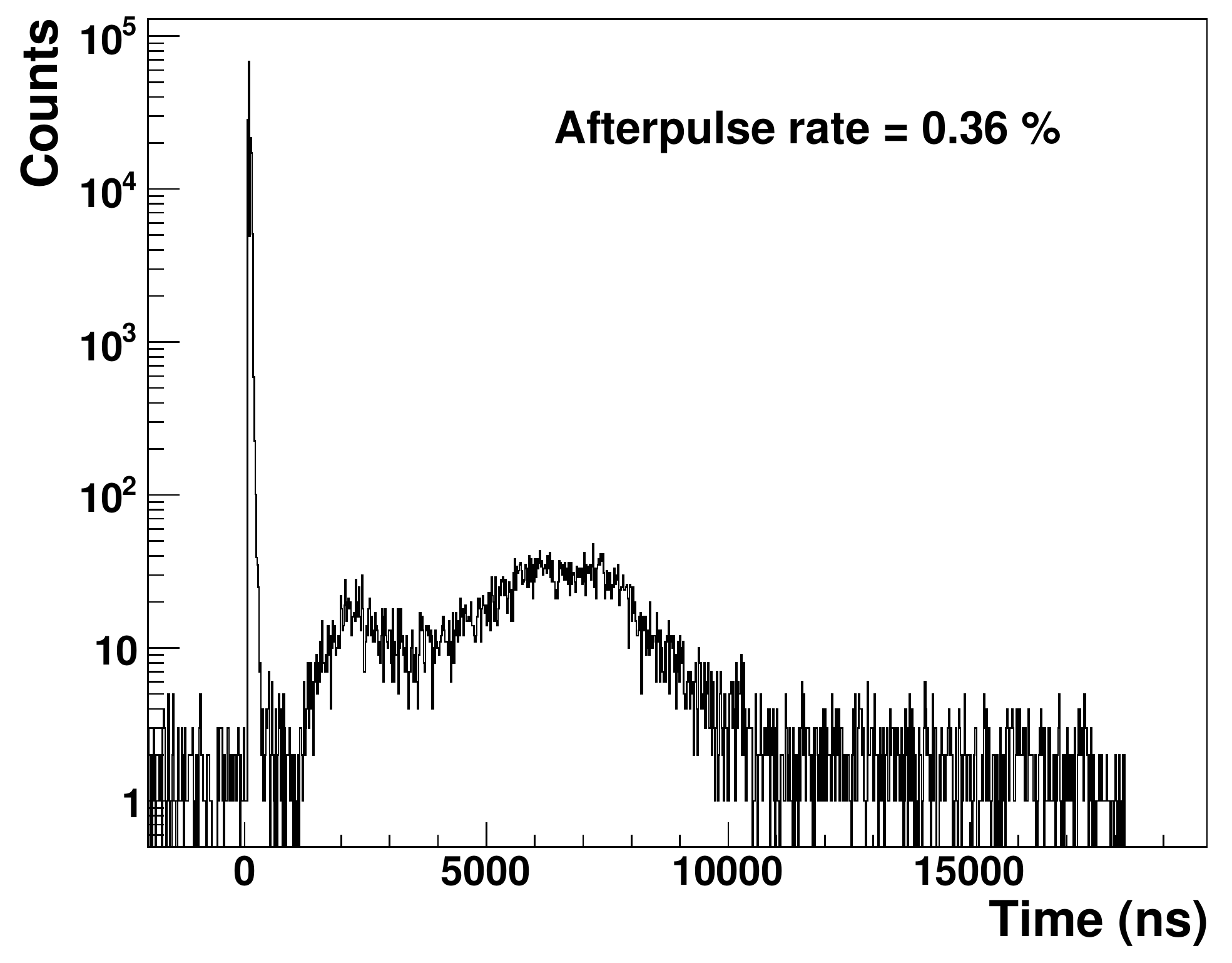}
\caption{The afterpulse distribution. The main pulse sits at around zero. The afterpulse rate (production probability) in the time range between 100 ns and 10 $\mu$s is 0.36\%.} \label{fig:ap}
\end{center}
\end{figure}

Afterpulses are mainly caused by positive ions generated by the ionization of residual gas after the impact of electrons from the photocathode and dynodes. The PMT signal was discriminated with a threshold of 1/3 PE. The afterpulse measurement was performed using a multi-hit TDC~\cite{Zhao:2016zht,Jiang:2020rdv}. Fig.~\ref{fig:ap} shows the afterpulse distribution. The main pulse sits at around zero. We define the afterpulse rate as the probability for generating an afterpulse by a single photoelectron in the main pulse~\cite{Zhao:2016zht}. The afterpulse rate between 100 ns and 10 $\mu$s is 0.36\%.

\subsection{Performances of the other three CR365-02-2 PMTs}
To validate the base design, the other three CR365-02-2 PMTs were also tested. The test results are summarized in Table~\ref{tab:4pmt}. The results confirm the conformity of CR365-02-2 and the base design to the requirements of the LHAASO-MD experiment.
\begin{table*}
\centering
  \begin{tabular}{c|c|c|c|c}
    PMT ID  & SA0190 & SA0217 & SA0221 & SA0225 \\\hline
    Working HV (V) &  1260 & 1284 & 1363 & 1288 \\
    P/V & 3.5 & 2.9 & 3.3 & 2.6 \\
    $\sigma_1/\mu_1$ & 47.7\% & 49.4\% & 47.3\% & 50.6\% \\
    Single muon resolution & 13.2\% & 13.3\% & 13.2\% & 13.4\% \\
    A/D & 151.7 & 146.0 & 120.5 & 149.0 \\
    Anode dynamic range (mA) & 34.4 & 32.0 & 31.1 & 36.0 \\
    DY7 dynamic range (mA) & 1775 & 1869 & 1627 & 1773 \\
    Dark count rate (cps) & 505 & 416 & 515 & 472 \\
    Afterpulse rate & 0.36\% & 0.58\% & 0.66\% & 0.58\% \\
\end{tabular}
\caption{Test results of four CR365-02-2 PMTs.}\label{tab:4pmt}
\end{table*}

\section{Conclusions}\label{conclusions}
A high dynamic range base for the CR365-02-2 8-inch PMT has been designed for the LHAASO-MD detector. It relies only on resistors and capacitors. The base comprises two outputs: the anode and DY7. This design ensures a good SPE resolution, with P/V great than 2. The charge resolution for single muons is below 25\%. The dynamic ranges (non-linearity within 5\%) are larger than 25 mA and 1600 mA (equivalent anode peak current) for the anode and DY7 outputs, respectively. The anode-to-dynode amplitude ratio is below 160 to ensure enough overlaps between the two outputs. The test results of 4 PMTs confirm the conformity of CR365-02-2 and the base design to the requirements of the LHAASO-MD detector.

\section*{Acknowledgments}
The research presented in this paper received strong support from the LHAASO collaboration and National Natural Science Foundation of China (No. 11675172, 11775217). We want to thank Profs. Zhen Cao, Huihai He, Xiangdong Sheng, Gang Xiao, Xiong Zuo, Cong Li, and other members of the LHAASO Collaboration for their valuable support and suggestions. 

%\section*{References}

\bibliography{mybibfile}

\end{document}